\documentclass[prc,showpacs,showkeys,nofootinbib,superscriptaddress]{revtex4}
\usepackage{graphicx}
\usepackage{dcolumn}
\usepackage{subfigure}
\usepackage{epsfig}
\usepackage{bm}
\usepackage{color}
\newcommand{\eq}{\begin{eqnarray}}
\newcommand{\en}{\end{eqnarray}}

\newcommand{\ba}[1]{\begin{eqnarray} \label{(#1)}}
\newcommand{\ea}{\end{eqnarray}}

\newcommand{\newc}{\newcommand}
\newc{\lra}{\leftrightarrow}
\newc{\beq}{\begin{equation}}
\newc{\eeq}{\end{equation}}
\newc{\barr}{\begin{eqnarray}}
\newc{\earr}{\end{eqnarray}}
  \def\vbf{\mbox{\boldmath $\upsilon$}}
    \def\sbf{\mbox{\boldmath $\sigma$}}
    \newcommand{\tritium}{^3{\rm H}{}}
\newcommand{\Hethree}{^3{\rm He}{}}
\begin{document}

\topmargin -0.50in
\title {Heavy sterile neutrino in dark matter searches}
\author{Paraskevi C. Divari }

\affiliation{ {\it Department of Physical Sciences and
Applications,Hellenic Army Academy, Vari 16673, Attica, Greece}
\footnote{pdivari@gmail.com}}
\author{ John D. Vergados}
\affiliation{ {\it TEI of Western Macedonia, Kozani, Gr 501 00,
Greece } \footnote{Permanent address, University of Ioannina,
Ioannina, Gr 451 10, Greece}}

%

\begin{abstract}
Sterile neutrinos are possible dark matter candidates. We examine
here possible detection mechanisms, assuming that the neutrino has
a mass of about 50 keV and couples to the ordinary neutrino.
 Even though this neutrino is quite heavy, it
is non relativistic with a maximum kinetic energy of 0.1 eV. Thus
new experimental techniques are required for its detection. We
estimate the expected event rate in the following cases: i)
Measure electron recoils in the case of materials with very low
electron binding. ii) Low temperature crystal bolometers. iii)
Spin induced atomic excitations at very low temperatures, leading
to a characteristic photon spectrum. iv) Observation of resonances
in antineutrino absorption by a nucleus undergoing electron
capture.
 v) Neutrino induced electron events beyond the end point energy of beta
 decaying systems, e.g. in the tritium decay studied by KATRIN.
\end{abstract}


\date{\today}

\maketitle
\section{Introduction}
There exist evidence for  existence of dark matter in almost  all
scales, from the dwarf galaxies, galaxies and cluster of galaxies,
the most important being the observed rotational curves in the
galactic halos, see e.g. the review \cite{UK01}. Furthermore
cosmological observations have provided plenty of additional
evidence ,  especially the recent
WMAP  \cite{WMAP06} and Planck \cite{PlanckCP15} data.

 In spite of this plethora of evidence, it is clearly  essential to directly
detect such matter in the laboratory in order to unravel its
nature.
 At present there exist many such candidates, called Weakly Interacting Massive Particles (WIMPs). Some examples are the
LSP (Lightest Supersymmetric Particle)
\cite{ref2a,ref2b,ref2c,ref2d,ELLROSZ,ELLFOR}, technibaryon
\cite{Nussinov92,GKS06}, mirror matter\cite{FLV72,Foot11} and
Kaluza-Klein models with universal extra
dimensions\cite{ST02a,OikVerMou}. Among other things these models
predict an interaction of dark matter with ordinary matter via the
exchange of a scalar particle, which leads to a spin independent
interaction (SI), or vector boson interaction, and therefore to a
spin dependent (SD) nucleon cross section.

Since the WIMP's are  expected to be
extremely non-relativistic, with average kinetic energy $\langle T\rangle  \approx
50 \ {\rm keV} (m_{\mbox{{\tiny WIMP}}}/ 100 \ {\rm GeV} )$, they are
not likely to excite the nucleus, even if they are quite massive $m_{\mbox{{\tiny WIMP}}} > 100$ GeV.
Therefore they can be directly detected mainly via the recoiling of a nucleus, first proposed more than 30 years ago \cite{GOODWIT}.
There exists a plethora of direct dark matter experiments  with the task of detecting  WIMP event rates
 for a variety of targets such as those employed in XENON10 \cite{XENON10}, XENON100 \cite{XENON100.11},
  XMASS \cite{XMASS09}, ZEPLIN \cite{ZEPLIN11}, PANDA-X \cite{PANDAX11}, LUX \cite{LUX11}, CDMS \cite{CDMS05},
  CoGENT \cite{CoGeNT11}, EDELWEISS \cite{EDELWEISS11}, DAMA \cite{DAMA1,DAMA11}, KIMS \cite{KIMS07}
  and PICASSO \cite{PICASSO09,PICASSO11}. These consider dark matter candidates in the multi GeV region.

Recently, however, an important  dark matter particle candidate of
the Fermion variety in the mass range of   10-100 keV, obtained
from galactic observables, has arisen
\cite{Mavromatos16,RAR16,ARR16}. This scenario  produces basically
the same behavior in the power spectrum (down to Mpc scales) with
that of standard $\Lambda$CDM cosmologies, by providing the
expected large-scale structure \cite{Boyarskyb09}. In addition, it
is not too warm, i.e. the masses involved are larger than $m= 1- 3
$keV to be  in conflict  with the current Ly$\alpha$ forest
constraints \cite{Boyarskyc09} and the number of Milky Way
satellites \cite{Tollerud08} , as in standard $\Lambda$WDM
cosmologies. In fact an interesting  viable candidate has been
suggested, namely a  sterile neutrino in the mass region of 48
-300 keV
\cite{Mavromatos16,RAR16,ARR16,camp16,YFLi1,YFLi2,Arcadi16,WeiLiao},
but most likely around 50 keV. For a recent review, involving a wider range of masses, see the white
paper \cite{WhitePaper16}.

The existence of light sterile neutrinos had already  been
introduced to explain some experimental anomalies like those
claimed    in the short baseline LSND and MiniBooNE experiments
\cite{LSND01,MiniBooNE01,KatCon01}, the reactor neutrino deficit
\cite{Mueller11} and the Gallium anomaly
\cite{Abdurashitov11,KopMalSch13}, with possible interpretations
discussed, e.g., in Refs \cite{GLLL13,Mention11} as well as in
\cite{Asaka05,Kusenko09} for sterile neutrinos in the keV region.
The existence of light neutrinos can be expected in an extended
see-saw mechanism  involving a suitable neutrino mass matrix
containing a number of neutrino singlets not all of which being
very heavy. In such models is not difficult to generate more than
one sterile neutrino, which can couple to the standard neutrinos
\cite{HeurtTer16}. As it has already mention, however, the
explanation of cosmological observations require sterile neutrinos
in the 50 keV region, which can be achieved in various models
\cite{Mavromatos16,MavPil12}.

In the present paper we will examine possible direct detection
possibilities for the direct detection of these  sterile
neutrinos. Even though these neutrinos are quite heavy, their
detection is not easy. Since like all dark matters candidates move
in our galaxies with not relativistic velocities, with average
value  about $10^{-3}$c and with energies about 0.05 eV, not all
of which can be deposited in the detectors. Therefore the standard
detection techniques employed in the standard dark matter
experiments like those mentioned above are not applicable in this
case. Furthermore, the size of the mixing parameter of sterile
neutrinos with ordinary neutrinos is crucial for detecting sterile
neutrinos. Thus our results concerning  the  expected  event rates
will be given in terms of this parameter.

The paper is organized as follows. In section II we study the
option on neutrino electron scattering.  In section III  we
consider the case of low temperature bolometers. In section IV the
possibility of neutrino induced atomic excitations is explored. In
section V we will consider the antineutrino absorption on nuclei,
which normally undergo electron capture, and finally  in section
VI   the modification of the end point electron energy in beta
decay, e.g. in the KATRIN experiment~\cite{katrin} is discussed.
In section VII, we summarize our conclusions.

\section{The neutrino-electron scattering}

The sterile neutrino as dark matter candidate can be treated in
the framework of the usual dark matter searches for light WIMPs
except that its mass is very small. Its velocity follows a
Maxwell-Boltzmann (MB) distribution with a characteristic velocity
about $10^{-3}$c. Since the sterile neutrino couples to the
ordinary electron neutrino it can be detected in neutrino electron
scattering experiments with the advantage that the
neutrino-electron cross section is very well known. Both the
neutrino and the electron can be treated as non relativistic
particles. Furthermore we will assume that the electrons are free,
since the WIMP energy is not adequate to ionize an the atom. Thus
the differential cross section is given by: \beq
d\sigma=\frac{1}{\upsilon} C^2_{\nu}(g_V^2+g^2_A)\left
(\frac{G_F}{2 \sqrt{2}}\right )^2\frac{d^3{\bf p}'_{\nu}}{(2
\pi)^3} \frac{d^3{\bf p}_{e}}{(2 \pi)^3} (2 \pi)^4 \delta({\bf
p}_{\nu}-{\bf p}'_{\nu}-{\bf p}_e)\delta(\frac{p^2_{\nu}}
{2m_{\nu}}-\frac{(p')^2_{\nu}}{2m_{\nu}}-\frac{p^2_{e}}{2m_{e}})
\eeq where $C^2_{\nu}$ is the square of the mixing of the sterile
neutrino with the standard electron neutrino $\nu_e$ and
$G_F=Gcos\theta_c$ where $G =1.1664\times 10^{-5}$$ GeV^{-2}$
denotes the Fermi weak coupling constant and $\theta_c\simeq 13^o$
is the Cabibbo angle~\cite{Beringer:2012zz}. The integration over
the outgoing neutrino momentum is trivial due to the momentum
$\delta$ function yielding: \beq d\sigma=\frac{1}{\upsilon}
C^2_{\nu}(g_V^2+g^2_A)\left (\frac{G_F}{2 \sqrt{2}}\right
)^2\frac{1}{(2 \pi)^2} d^3{\bf p}_{e} \delta(p_e \upsilon
\xi-\frac{p^2_{e}}{2\mu_{r}}) \label{Eq:WIMPcross} \eeq where
$\xi=\hat{p}_e.\hat{p}_{\nu}$, $0\le\xi\le 1$, $\upsilon$ the WIMP
velocity and $\mu_{r}$ is the WIMP-electron reduced mass, $\mu_{r}
\approx m_{\nu}$. The electron energy $T$ is given by: \beq
T=\frac{p^2_e}{2 m_e}= 2 \frac{m^2_{\nu}}{m_e}  \left (\upsilon
\xi\right )^2 \Rightarrow 0\leq T\leq 2 \frac{m^2_{\nu}}{m_e}
\upsilon^2_{esc} \label{Eq.TvsV} \eeq
 where $\upsilon_{esc}$ is the maximum WIMP
velocity (escape velocity). Integrating Eq. (\ref{Eq:WIMPcross})
over the angles, using the $\delta$ function for the  $\xi$
integration we  obtain: \barr
d\sigma&=&C^2_{\nu}\frac{1}{\upsilon} (g_V^2+g^2_A)\left
(\frac{G_F}{2 \sqrt{2}}\right )^2\frac{1}{2 \pi}  p^2_{e}d
p_e\frac{1}{|p_e \upsilon|}\Rightarrow \nonumber\\
d\sigma&=&C^2_{\nu}\frac{1}{\upsilon^2} (g_V^2+g^2_A)\left
(\frac{G_F}{2 \sqrt{2}}\right )^2\frac{1}{ 2\pi}  m_e dT \earr

\begin{figure}
  \begin{center}
\includegraphics[scale=1,width=8cm]{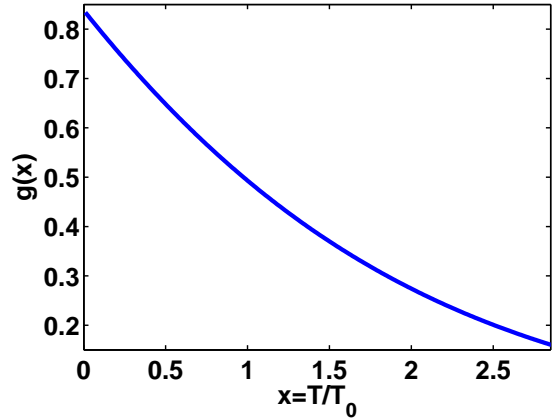}
 \caption{ The shape of the spectrum of the emitted electrons in sterile neutrino-electron scattering}
 \label{fig:nuescat}
 \end{center}
  \end{figure}

We are now in a position to fold it the velocity distribution
assuming it to be MB with respect to the galactic center
 \beq
f(\upsilon')=\frac{1}{\left ( \sqrt{\pi}\upsilon_0\right
)^3}e^{-\left (\upsilon'/\upsilon_0\right )^2} \eeq In the local
frame, assuming that the sun moves around the center of the galaxy
with velocity $\upsilon_0=220\mbox{km/s}$, $\vbf'= \vbf+\upsilon_0
\hat{z}$, we obtain: \beq f_{\ell}(y,\xi)=\frac{1}{\left
(\sqrt{\pi}\upsilon_0\right )^3}e^{- (1+y^2+2 y \xi)}, \quad
y=\frac{\upsilon}{\upsilon_0}, \eeq where $\xi$ is now the cosine
of the angle between the WIMP velocity $\vbf$  and the direction
of the sun's motion. Eventually we will need the flux so we
multiply with the velocity $\upsilon$ before we integrate over the
velocity. The limits of integration are between $\upsilon_{min}$
and $\upsilon_{esc}$. The velocity is given via Eq.
(\ref{Eq.TvsV}), namely: \beq \upsilon=\frac{\sqrt{2 m_e T}}{2
m_{\nu} \xi}\Rightarrow \upsilon_{min}=\frac{\sqrt{2 m_e
T}}{2m_{\nu}} \eeq We find it convenient to express the kinetic
energy $T$ in units of $T_0=2(m^2_{\nu}/m_e)\upsilon^2_0$. Then
$$ y_{min}=\sqrt{x},\quad  x=\frac{T}{T_0}.$$
Thus
$$
\langle \upsilon \frac{d\sigma}{dT}\rangle = \frac{1}{\upsilon_0}
\frac{1}{T_0}\frac{m_e}{16 \pi}C^2_{\nu}
G^2_F(g_V^2+g^2_A)\int_{\sqrt{x}}^{y_{esc}} dy y
\frac{2}{\sqrt{\pi}}e^{-(1+y^2)}\int_{-1}^1 d \xi e^{-2 y \xi}
$$
These integrals can be done analytically to yield
\begin{eqnarray}
&\displaystyle \langle \upsilon \frac{d\sigma}{dT}\rangle
= \frac{1}{\upsilon_0}\frac{m_e}{16 \pi}
 C^2_{\nu}G^2_F(g_V^2+g^2_A)   g(x), \\
\nonumber \\
 & g(x)=\frac{1}{2}
   \left(\text{erf}\left(1-\sqrt{x}\right)+\text{erf}\left(\sqrt{x}+1
   \right)+\text{erfc}(1-y_{esc})+\text{erfc}(y_{esc}+1)-2\right)\nonumber
\end{eqnarray} where $\text{erf}$ is the error function and
$\text{erfc}(x)$ is its complement. The function $g(x)$
characterizes the spectrum of the emitted electrons and is
exhibited in Fig. \ref{fig:nuescat}
 and it is without any particular structure, which is the case in most WIMP searches.
%
For a 50 keV sterile neutrino we find that:
$$T_0=2 \left(\frac{m_{\nu}}{m_e}\right )^2 \left(\frac{2.2}{3}\right)^2 10^{-6} m_e c^2\approx 5.0 \times 10^{-3}\mbox{eV}$$
$$T_{max}=T_0 y_{esc}^2=5\times 10^{-3} 2.84^2\approx 0.04 \mbox{eV}$$
$$\langle T \rangle=1.6T_0=8.0\times 10^{-3} \mbox{eV} $$

Now $dT=T_0dx$. Thus
    \beq
\frac{\langle \upsilon
\sigma\rangle}{\upsilon_0}=\frac{1}{\upsilon_0^2}\frac{m_e T_0}{16
\pi} C^2_{\nu}G^2_F(g_V^2+g^2_A) \int_0^{y_{esc}^2} dx g(x)= 1.43
 \frac{m_{\nu}^2}{8 \pi} C^2_{\nu}G^2_F(g_V^2+g^2_A)
 \eeq
 where
 $$\int_0^{y_{esc}^2} dx g(x)= 1.43$$

It is clear that with this amount of energy transferred to the electron it is not
possible to eject an electron out of the atom. One therefore must use special materials such that the electrons are loosely bound.
 It has recently been suggested that it  is possible to detect even very light WIMPS, much lighter than the electron, utilizing
 Fermi-degenerate materials like superconductors\cite{HPZ15}. In this case the energy required is essentially the gap energy of
 about $1.5 kT_c$, which is in the meV region, i.e the electrons are essentially
 free. In what follows, we assume the values
\begin{eqnarray}
&g_A=1, \quad g_V=1+4\sin^2\theta_W=1.92,
\nonumber \\
 & G^2_F=5.02\times 10^{-44}\hspace{2pt}  cm^2/MeV^2
\end{eqnarray}
while   $C^2_{\nu}$ is taken as a parameter and will be discussed
in section VII.
 Thus we obtain:
\begin{eqnarray}
&\displaystyle\frac{\langle \upsilon \sigma\rangle}{\upsilon_0}=
 &=3.47\times
 10^{-47}C^2_{\nu}   \quad \mbox{cm}^2
\end{eqnarray}
The
neutrino particle density is
$$N_{\nu}=\frac{\rho}{m_{\nu}} =\frac{0.3 \mbox{GeV/cm}^{3}}{50 \times 10^{-6}\mbox{GeV}}=6\times 10^{3}\mbox{cm}^{-3}$$
while   the neutrino flux
$$\Phi_{\nu}=\frac{\rho}{m_{\nu}}\upsilon_0=1.32\times 10^{11}cm^{-2}s^{-1}$$
where $\rho=0.3~ \mbox{GeV/cm}^{3}$ being the dark matter density.
 Assuming that the number of electron pairs in the
target is $2\times N_A=2\times 10^{24}$ we find  that the number
of events per year is
\begin{eqnarray}
 \Phi_{\nu}  \frac{\langle \upsilon \sigma\rangle}{\upsilon_0}
 2\times N_A&=2.89\times 10^{-4} C^2_{\nu}\mbox{y}^{-1}
\label{Eq:recoilrate}
\end{eqnarray}

The authors of \cite{HPZ15} are perhaps aware of the fact that the
average energy for very light WIMPS is small and, as we have seen
above,  a small portion  of it is transferred to their system .
With their bolometer   detector these authors  probably have a way
to circumvent the fact that  a small amount  of energy  will be
deposited, about 0.4 eV in a year for $N_A\approx 10^{24}$.
Perhaps they may manage to accumulate a larger number of loosely
bound electrons  in their target.

\section{Sterile neutrino detection via low temperature bolometers}

Another possibility is to use bolometers, like the CUORE detector
exploiting Low Temperature Specific Heat of Crystalline
$^{130}$TeO$_2$ at low  temperatures. The energy of the WIMP will
now be deposited in the crystal, after its interaction with the
nuclei via Z-exchange. In this
 case the Fermi component of interaction with neutrons is coherent, while that of the protons is negligible. Thus the matrix element becomes:
\beq
\mbox{ME}=\frac{G_F}{2 \sqrt{2}}N g_V,\,N=\mbox{number of neutrons in the nucleus}
\eeq
 A detailed analysis  of the frequencies the  $^{130}$TeO$_2$ can  be found \cite{TeNorMod06}.
 The analysis involved  crystalline phases of tellurium dioxide: paratellurite $\alpha$-TeO$_2$,
 tellurite $\beta$-TeO$_2$ and the new
phase γ-TeO2, recently identified experimentally. Calculated
Raman and IR spectra are in good agreement with available
experimental data. The vibrational spectra of $\alpha$ and
$\beta$-TeO$_2$ can be interpreted in terms of vibrations of
TeO$_2$ molecular units. The $\alpha$-TeO$_2$ modes are associated
with the symmetry $D_4$ or 422, which has 5 irreducible
representations, two 1-dimensional of the antisymmetric type
indicated by $A1$ and $A_2$,  two 1-dimensional of the symmetric
type  $B1$ and $B_2$ and one 2-dimensional, usually indicated by
$E$. They all have been tabulated in Ref.~\cite{TeNorMod06}. Those
that can be excited must be below the Debye frequency which has
been determined \cite{CUORE01} and found  to be quite low:
$$T_D=(232\pm 7)\, ^{0}\mbox{K}\Rightarrow \omega_D=0.024\mbox{eV}$$
This frequency is smaller than the maximum sterile neutrino energy
estimated to be $T_{max}=0.11$ eV. Those frequency modes of
interest to us are given in Table \ref{tab:modes}. The
differential  cross section is, therefore given by
\begin{table}
\begin{center}
\caption{The frequency modes below the Debye temperature for
$\alpha$-TeO$_2$ obtained from table VIII of ref.\cite{TeNorMod06}
(for notation see text).} \label{tab:modes}
$$
\begin{array}{c|cccccccc}
\hline\\
 \nu_i=\frac{\omega_i}{2 \pi}\left(\mbox{cm}^{-1}\right)&52 & 124 & 128 & 152 & 157 & 176 &
   177 & 179 \\
\mbox{symmetry}& B_1 & E & B_1 & A_1 & B_2 & A_2
   & E & B_1 \\
\omega_i\mbox{(eV)}& 0.006 & 0.015 & 0.016 & 0.019 &
   0.019 & 0.022 & 0.022 & 0.022 \\
 N_i&16 & 6 & 6 & 5& 5 & 4 & 4 & 4 \\
 E_{max}(i)\mbox{(meV)}& 106& 100& 102& 103& 107&98
&99&100 \\
\end{array}
$$
\end{center}
\end{table}

\beq
d\sigma=\frac{1}{\upsilon} C^2_{\nu} N^2 (g_V^2)
\left (\frac{G_F}{2 \sqrt{2}}\right )^2\sum_{k=1}^8\sum_{n_i=0}^{N_k}\frac{d^3{\bf p}'_{\nu}}{(2 \pi)^3}
 \frac{d^3{\bf q}}{(2 \pi)^3} (2 \pi)^4 F^2({\bf q}^2)\delta({\bf p}_{\nu}-{\bf p}'_{\nu}-{\bf q})\delta
 \left (\frac{p^2_{\nu}}{2m_{\nu}}-\frac{(p')^2_{\nu}}{2m_{\nu}}-(n_i+\frac{1}{2})\omega_k\right )
\eeq
where $N_i$ will be specified below and ${\bf q}$ the momentum transferred to the nucleus.
The momentum transfer is small and the form factor $F^2({\bf q}^2)$ can be neglected.

In deriving this formula we tacitly assumed a coherent interaction between
the WIMP and several nuclei, thus creating  a collective
excitation of the crystal, i.e. a phonon or few phonons. This of course is a good approximation
provided that the energy transferred is small, of a few tens of meV. We see from table \ref{tab:modes}
that the maximum allowed energy is small, around 100 meV. We find that, if we restrict the maximum allowed
energy by a factor of 2, the obtained results are reduced only by a factor of about $10\%$. We may thus
 assume that this approximation is good.

\begin{figure}
  \begin{center}
\includegraphics[scale=1,width=8cm]{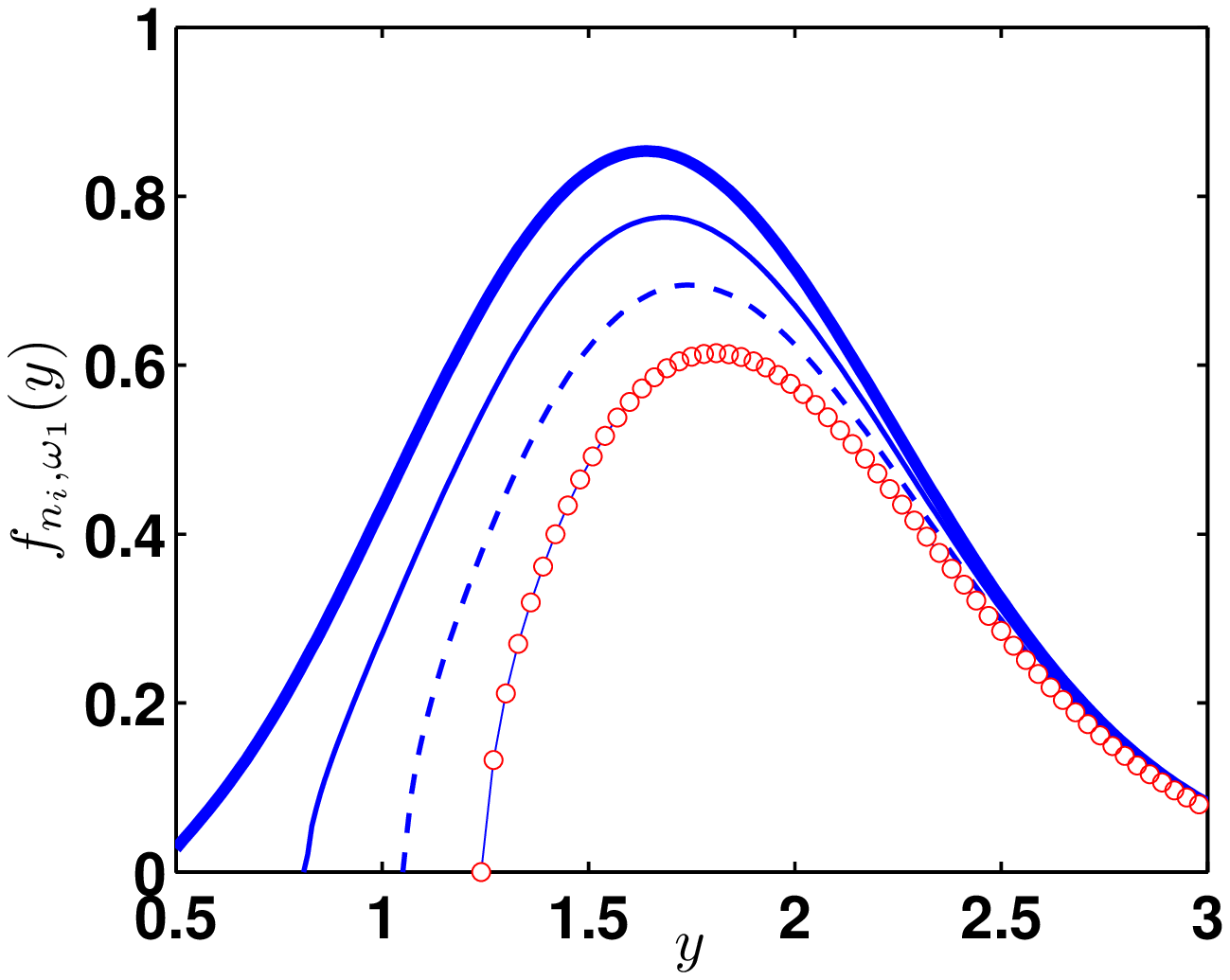}
\includegraphics[scale=1,width=8cm]{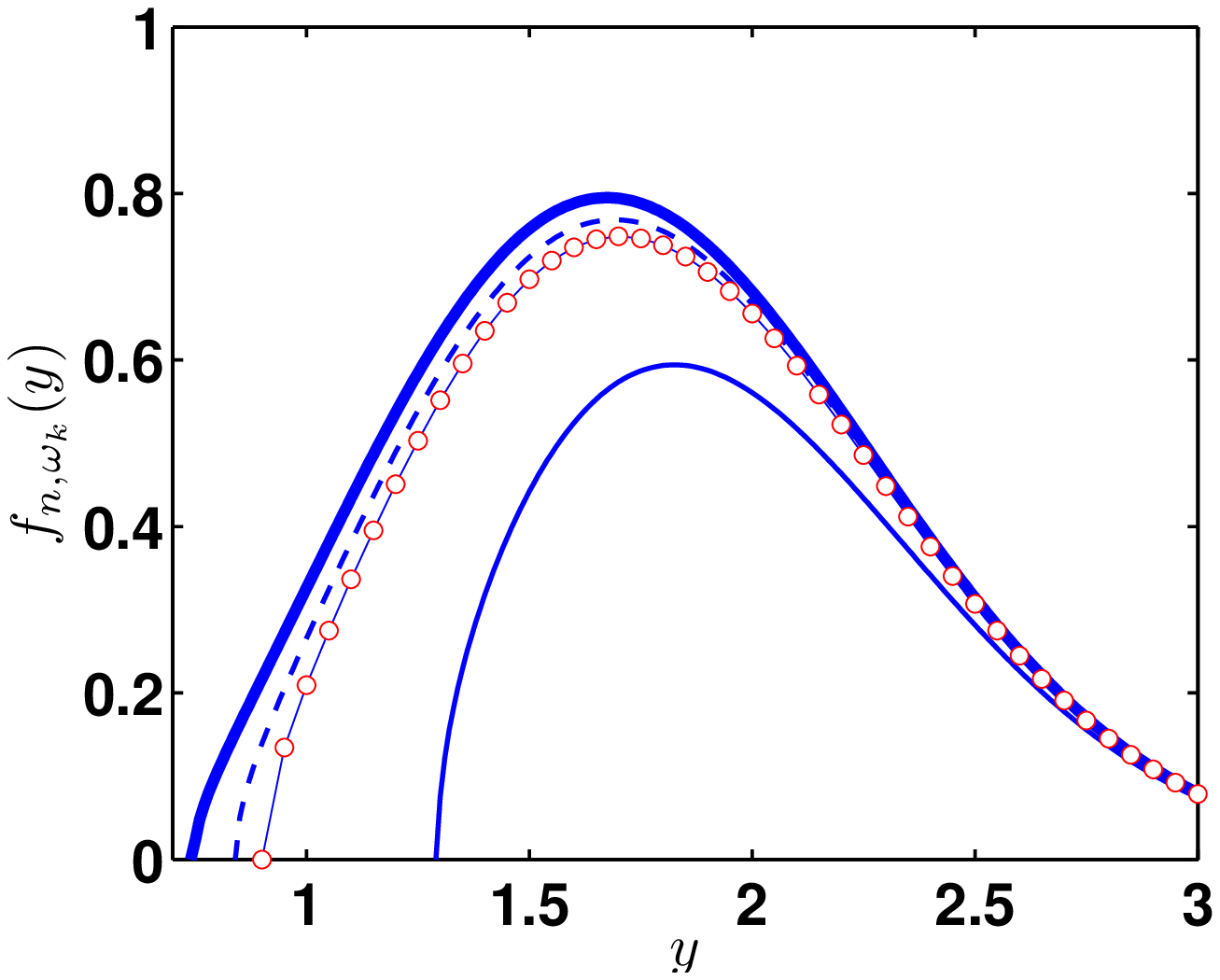}
\\
 \caption{ (a) The function $f_{n,\omega_1}(y)$, exhibited as a function of $y$,
 associated with the mode $\nu_1=52$cm$^{-1}$ for $n=0,1,2,3$ increasing downwards.(b)
   The functions $f_{n,\omega_2}(y)$ associated with $\nu_2=124$cm$^{-1}$ for $n=0,1$
   and $f_{0,\omega_4}$ for $\nu_4=157$cm$^{-1}$, $n=0$ and $f_{0,\omega_6}$ for $\nu_6=176$cm$^{-1}$, $n=0$,
   exhibited  as a function of $y$, for thick solid, solid,   dashed and   dotted lines respectively.
   For definitions see text. }
 \label{fig:cuore}
 \end{center}
  \end{figure}

 Integrating over the nuclear momentum we get
\beq d\sigma=\frac{1}{\upsilon} C^2_{\nu} N^2 (g_V^2) \left
(\frac{G_F}{2 \sqrt{2}}\right
)^2\frac{1}{(2\pi)^2}\sum_{k=1}^8\sum_{n_i=0}^{N_k}d^3{\bf
p}'_{\nu}\delta \left
(\frac{p^2_{\nu}}{2m_{\nu}}-\frac{(p')^2_{\nu}}{2m_{\nu}}-(n_i+\frac{1}{2})\omega_k\right
) \eeq \beq d\sigma=\frac{1}{\upsilon} C^2_{\nu} N^2 (g_V^2)\left
(\frac{G_F}{2 \sqrt{2}}\right )^2
\frac{1}{(2\pi)^2}\sum_{k=1}^8\sum_{n_i=0}^{N_k}d^3{\bf
p}'_{\nu}\delta \left
(\frac{p^2_{\nu}}{2m_{\nu}}-\frac{(p')^2_{\nu}}{2m_{\nu}}-(n_i+\frac{1}{2})\omega_k\right
) \eeq performing the integration using the $\delta$ function we
get \beq \sigma=\frac{1}{\upsilon} C^2_{\nu} N^2 (g_V^2) \left
(\frac{G_F}{2 \sqrt{2}}\right )^2\frac{1}{\pi} m_{\nu}\sqrt{2
m_{\nu}}\sum_{k=1}^8\sum_{n_i=1}^{N_k}
\sqrt{E_{\nu}-(n_i+\frac{1}{2})\omega_k} \eeq \beq
\sigma=\frac{\upsilon_0}{\upsilon} C^2_{\nu} N^2 (g_V^2)\left
(\frac{G_F}{2 \sqrt{2}}\right )^2 \frac{1}{\pi}
m^2_{\nu}\sum_{k=1}^8\sum_{n_i=0}^{N_k}\sqrt{y^2-\frac{(n_i+\frac{1}{2})\omega_k}{T_1}}
\eeq where
$T_1=\frac{1}{2}m_{\nu}\upsilon_0^2,\,y=\frac{\upsilon}{\upsilon_0}$\\Folding
with the velocity distribution we obtain \barr \langle
\upsilon\sigma\rangle&=& \upsilon_0 C^2_{\nu} N^2 (g_V^2)\left
(\frac{G_F}{2 \sqrt{2}}\right )^2 \frac{1}{\pi}
m^2_{\nu}\sum^8_{k=1}\sum_{n_i=0}^{N_k}
I_{n_i,\omega_k},\nonumber\\I_{n_i,\omega_k}
&=&\int_{y_{min}}^{y_{esc}}dy
f_{n_i,\omega_k}(y),\,f_{n_i,\omega_k}(y)=
\sqrt{y^2-\frac{(n_i+\frac{1}{2})\omega_k}{T_1}}y
e^{-1-y^2}\sinh{2 y},\\
y_{min}&=&\sqrt{\frac{(n_i+\frac{1}{2})\omega_k}{T_1}}\nonumber
\earr We see that we have the constraint imposed by the available
energy, namely:
$$N_k=IP \left [\frac{y^2_{esc}T_1}{\omega_k}-\frac{1}{2}  \right ]$$ where
$IP[x]=$integer part of $x$.  We thus find the $N_k$ listed in table \ref{tab:modes}.
The functions $f_{n_i,\omega_k}(y)$ are exhibited in fig. \ref{fig:cuore}.
The relevant  integrals are  $I_n(\omega_1)=(1.170, 0.972, 0.785,
0.621)$ for $n=0,1,2,3$, $I_{n}(\omega_2)=(1.032, 0.609)$, for
$n=0,1$, $I_{n}(\omega_3)=(1.025, 0.592)$, for $n=0,1$ and
$I_0(\omega_k)= (0.979, 0.970, 0.934, 0.932, 0.929)$ for
$k=4,\cdots,8$. Thus  we obtain a total of 17.8. The event rate
takes with a target of mass $m_t$ takes the the form:
$$R=\Phi_{\nu}C^2_{\nu} N^2 (g_V^2)G^2_F\frac{1}{8\pi}\frac{m_t}{A m_p} m^2_{\nu}17.8$$
If we restrict the maximum allowed energy to half of that shown in
Table \ref{tab:modes} by
a factor of two, we obtain 15.7 instead of 17.8.\\
For a $^{130}$TeO$_2$ target (N=78) of 1 kg of mass get
$$R=1.7 \times 10^{-6}C^2_{\nu}\hspace{2pt}\mbox{ per kg-s}=51 C^2_{\nu} \hspace{2pt}\mbox{ per kg-y} $$
This is much larger than that obtained in the previous section,
mainly due to the neutron coherence arising from the Z-interaction
with the target (the number of scattering centers is approximately
the same $4.5 \times 10^{24})$. In the present case, however,
targets can be larger than 1 kg. Next we are going to examine
other mechanisms, which promise a better signature.

\section{Sterile neutrino detection via atomic excitations}

We are going to examine the interesting possibility of excitation of an atom
from a level $|j_1,m_1\rangle $ to a nearby   level $|j_2,m_2\rangle $ at energy $\Delta=E_2-E_1$,
 which has the same orbital structure. The  excitation energy has to be quite low, i.e:
\beq
\Delta \leq\frac{1}{2} m_{\nu}\upsilon^2_{esc}=\frac{1}{2}50 \times 10^3 2.84^2 \left (\frac{2.2}{3}\right )^2 10^{-6}=0.11 eV
\eeq
The target is selected so that the two levels  $|j_1,m_1\rangle$ and   $|j_2,m_2\rangle$ are closer han 0.11 eV.
This can arise  result from the splitting of an atomic level by the magnetic field so that they can be connected by
the spin operator. The lower one $|j_1,m_1\rangle$ is occupied by electrons but the higher one $|j_2,m_2\rangle$ is
completely empty at sufficiently low temperature. It can be populated only by exciting an electron to it from the
lower one by the oncoming sterile neutrino. The presence  of such an excitation is monitor by a tuned laser which
excites such an electron from    $|j_2,m_2\rangle$  to a higher state  $|j_3,m_3\rangle$, which cannot be reached
 in any other way, by observing its subsequent decay by emitting photons.

Since this is an one body transition the relevant matrix element takes the form:
\beq
|ME(j_1,m_1;j_2,m_2)|^2=g_V^2\delta_{j_1,j_2}\delta_{m_1,m_2}+g_A^2  \left (C_{\ell,j_1,m_1,j_2,m_2 }\right )^2
\eeq
 (in the case of the axial current we have  $g_A=1$ and we need evaluate the matrix
  element of $\sbf_{\nu}.\sbf_e$ and then square it and sum as well as average over the neutrino polarizations).
\beq C_{\ell,j_1,m_1,j_2,m_2 }=\langle n \ell j_2 m_2|\sbf|n\ell
j_1 m_1\rangle=\langle j_1\,m_1,1\,m_2-m_1|j_2\,m_2\rangle\sqrt{(2
j_1+1)3}\sqrt{2 \ell+1}\sqrt{6}\left \{
\begin{array}{ccc}\ell&\frac{1}{2}&j_1\\\ell&\frac{1}{2}&j_2
\\0&1&1 \end{array}\right \} \eeq expressed in terms of the
Glebsch-Gordan coefficient and the nine- j symbol. It is clear
that in the energy transfer of interest only the axial current can
contribute to excitation

 The cross section takes the form:
\beq
d\sigma=\frac{1}{\upsilon} C^2_{\nu}\left (\frac{G_F}{2 \sqrt{2}}\right )^2
|ME(j_1,m_1;j_2,m_2)|^2\frac{d^3{\bf p}'_{\nu}}{(2 \pi)^3} \frac{d^3{\bf p}_{A}}{(2 \pi)^3} (2 \pi)^4 \delta \left ({\bf p}_{\nu}-{\bf p}'_{\nu}-{\bf p}_A \right )\delta \left (E_{\nu}-\Delta-E'_{\nu} \right )
\eeq
Integrating over the atom recoil momentum, which has negligible effect on the energy,
and over the direction of the final neutrino momentum and energy via the $\delta$ function we obtain
\barr
\sigma&=&\frac{1}{\upsilon} C^2_{\nu}\left (\frac{G_F}{2 \sqrt{2}}\right )^2
|ME(j_1,m_1;j_2,m_2)|^2 \frac{1}{\pi}  \left (E_{\nu}-\Delta \right )\sqrt{2 \left (E_{\nu}-\Delta-m_{\nu} \right )m_{\nu}}\nonumber\\
&=&\frac{1}{\upsilon} C^2_{\nu}\left (\frac{G_F}{2 \sqrt{2}}\right
)^2 |ME(j_1,m_1;j_2,m_2)|^2 \frac{1}{\pi} m_{\nu}^2\sqrt{\frac{2
T_1}{m_{\nu}}} f\left (y,\frac{\Delta}{T_1}\right ) \earr

\barr f\left (y,\frac{\Delta}{T_1}\right )=\left(
y^2-\frac{\Delta}{T_1}\right )^{1/2},\,T_1=\frac{1}{2}m_{\nu}
\upsilon_0^2 \earr

where we have set $E-\Delta=m_{\nu}+T_1-\Delta \approx m_{\nu}$.

\begin{figure}
  \begin{center}
\includegraphics[scale=1,width=0.5\textwidth]{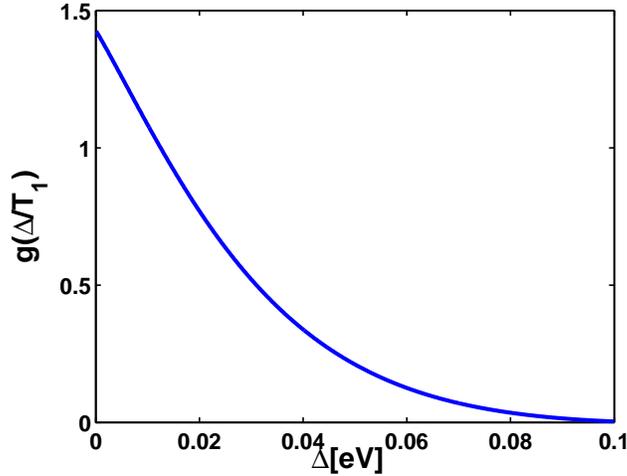}\\
 \caption{The function $g\left (\frac{\Delta}{T_1}\right )$
 for sterile neutrino scattering by an atom as a function of the excitation energy in eV.
 \label{fig:nueatom}}
 \end{center}
  \end{figure}

Folding the cross section with the velocity distribution from a
minimum $\sqrt{\frac{\Delta}{T_1}}$  to $y_{esc}$ we obtain
 \barr
\frac{\langle\upsilon \sigma\rangle}{\upsilon_0} &=&
C^2_{\nu}\left (\frac{G_F}{2 \sqrt{2}}\right )^2  m^2_{\nu}
|ME(j_1,m_1;j_2,m_2)|^2
\frac{1}{\pi}  g\left (\frac{\Delta}{T_1}\right )\nonumber\\
g\left (\frac{\Delta}{T_1}\right )
&=&\frac{2}{\sqrt{\pi}}\int_{\sqrt{\frac{\Delta}{T_1}}}^{y_{esc}}
dy y^2 \left( y^2-\frac{\Delta}{T_1}\right )^{1/2}e^{-(1+y^2)}
\frac{\sinh{2y}}{y} \earr Clearly the maximum excitation energy
that can be reached is $\Delta_{max}=2.84^2 T_0=0.108$eV. The
function $g\left (\frac{\Delta}{T_1}\right )$ is exhibited in Fig.
\ref{fig:nueatom}

Proceeding as in section II and noting that for small excitation
energy $g\left (\frac{\Delta}{T_1}\right )\approx 1.4$ we find :
    $$
    R=1.8 \times 10^{-2} C^2_{\nu} \hspace{2pt}\frac{1}{A} \left (C_{\ell,j_1,m_1,j_2,m_2 }\right )^2  \mbox{kg-y}.
$$
The expected rate will be smaller after the angular momentum factor $ C_{\ell,j_1,m_1,j_2,m_2 }$
is included (see appendix A). Anyway leaving aside this factor, which can only be determined after a
specific set of levels is selected, we see that the obtained rate is comparable to that expected from
electron recoil (see Eq. (\ref{Eq:recoilrate})). In fact for a target with $A=100$ we obtain:
$$ R=1.8 \times 10^{-4} C^2_{\nu} \hspace{2pt} \left (C_{\ell,j_1,m_1,j_2,m_2 }\right )^2  \mbox{kg-y}$$
This rate, however, decreases as the excitation energy increases
(see Fig. \ref{fig:nueatom}). In the present case, however, we
have two advantages.
\begin{itemize}
\item The characteristic signature of photons spectrum following
the de-excitation of of the level $|j_3,m_3\rangle$ mentioned
above. The photon energy can be changed if the target is put in a
magnetic field by a judicious choice of $|j_3,m_3\rangle$ \item
The target now can be much larger, since one can employ a solid at
very low temperatures. The ions of the crystal still exhibit
atomic structure. The electronic states probably won't carry all
the important quantum numbers as their corresponding neutral atoms. One may have to consider exotic atoms
(see appendix B) or  targets which
contain appropriate impurity atoms in a host crystal, e.g Chromium
in sapphire.
\end{itemize}
In spite of this it seems  very hard to detect such a process, since the expected counting rate is very low.

\section{Sterile neutrino capture by a nucleus undergoing  electron capture}

This is essentially the process: \beq \label{rec1}
\bar{\nu}+e_{b}+A(N,Z)\rightarrow A(N+1,Z-1)^{*} \eeq involving
the absorption  of a neutrino with the simultaneous capture of a
bound electron. It has already been studied \cite{JDVNov14} in
connection with the detection of the standard relic neutrinos. It
involves modern technological innovations like the Penning Trap
Mass Spectrometry (PT-MS) and the Micro-Calorimetry (MC). The
former should provide an answer to the question of accurately
 measuring the nuclear binding energies and how strong the resonance enhancement is expected, whereas the latter
 should analyze the bolometric spectrum in the tail of the peak corresponding to  L-capture to the excited state in
  order to observe the relic anti-neutrino events.   They also examined the suitability of $^{157}$Tb for relic
   antineutrino  detection via the resonant enhancement to  be considered by  the PT-MS and MC teams. In the
   present case the experimental constraints are expected to be less stringent since the sterile neutrino is much heavier.

Let us measure all energies from the ground state of the final
nucleus and assume that $\Delta$ is the mass difference of the two
neutral atoms. Let us consider a transition to the final state with
 energy $E_x$. The cross section for a
  neutrino\footnote{Here as well as in the following  we may write neutrino, but it is understood that we mean antineutrino}
   of given velocity $\upsilon$ and kinetic energy $E_{\nu}$ is given by:
\beq
\sigma(E_{\nu})= C^2_{\nu}\frac{1}{\upsilon} |ME(E_x)|^2_{\mbox{\tiny nuc}}
 \langle\phi_e\rangle^2\left (\frac{G_F}{2\sqrt{2}} \right )^2
 \frac{d^3p_A}{(2\pi)^3}(2 \pi)^4\delta({\bf p}_A-{\bf p}_{\nu}) \delta(E_{\nu}+m_{\nu}+\Delta-E_{x}-b)
\eeq
where ${\bf p}_A$ is the recoiling nucleus momentum. Integrating over the recoil momentum using the $\delta$ function we obtain
\beq
\sigma(E_{\nu})=C^2_{\nu}2 \pi \frac{1}{\upsilon} |ME(E_x)|^2_{\mbox{\tiny nuc}}
 \langle\phi_e\rangle^2\left (\frac{G_F}{2\sqrt{2}} \right )^2 \delta(E_{\nu}+m_{\nu}+\Delta-E_{x}-b),
\eeq
We note that, since the oncoming neutrino has a mass, the excited state must be
 higher than the highest excited state at $E'_x=\Delta-b$. Indicating by $\epsilon=E_x-E'_x$ the above equation can be written as
\beq
\sigma(E_{\nu})=C^2_{\nu}2 \pi \frac{1}{\upsilon} |ME(E_x)|^2_{\mbox{\tiny nuc}}
 \langle\phi_e\rangle^2\left (\frac{G_F}{2\sqrt{2}} \right )^2 \delta(E_{\nu}+m_{\nu}-\epsilon)
\eeq Folding it with the velocity distribution as above we obtain:
\beq \label{fol1} \langle \upsilon
\sigma(E_{\nu})\rangle=C^2_{\nu}2 \pi |ME(E_x)|^2_{\mbox{\tiny
nuc}} \langle\phi_e\rangle^2\left (\frac{G_F}{2\sqrt{2}} \right
)^2 \int_0^{y_{esc}} dy
 y^2 \frac{2}{\sqrt{\pi}}e^{-(1+y^2)}\frac{\sinh2y}{y}\delta\left (m_{\nu}+\frac{1}{2}m_{\nu}\upsilon_0^2 y^2-\epsilon\right )
\eeq or using the delta function

\begin{figure}
  \begin{center}
\includegraphics[scale=1,width=0.5\textwidth]{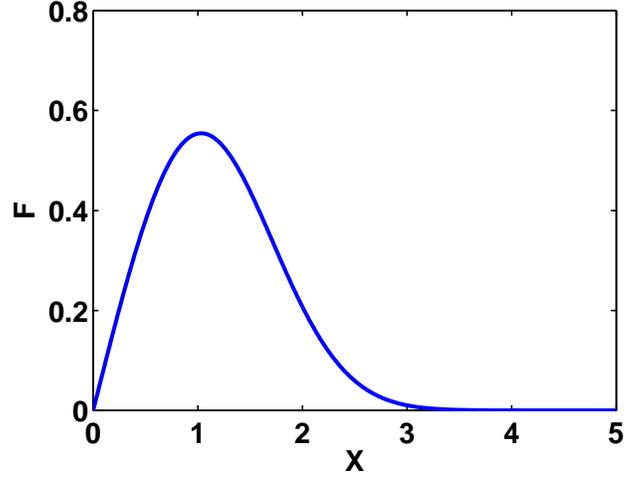}\\
 \caption{The cross section exhibits resonance behavior. Shown is the resonance
  properly normalized as a function of $X=\frac{1}{\upsilon_0}\sqrt{ 2\left (\frac{ \epsilon}{m_{\nu}}-1\right )}$.
  The width is $\Gamma=1.49 $ and the location of the maximum is at
 1.03
\label{fig:ecapwidth}}
\end{center}
 \end{figure}

\begin{figure}
\begin{center}
\includegraphics[scale=1,width=0.5\textwidth]{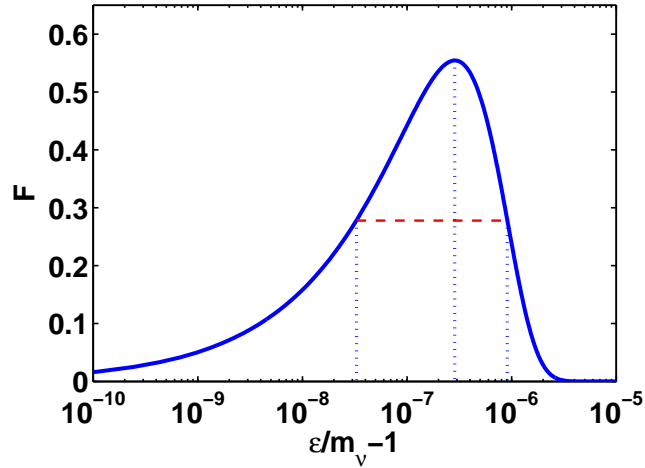}\\
 \caption{The cross section exhibits resonance behavior. Shown is $F(X)$
 as a function of $ \left (\frac{ \epsilon}{m_{\nu}}-1\right )$.
 \label{fig:ecapwidthE}}
\end{center}
\end{figure}

\barr \langle \upsilon \sigma(E_{\nu})\rangle&=& 2\pi C^2_{\nu}
|ME(E_x)|^2_{\mbox{\tiny nuc}} \frac{\langle\phi_e\rangle^2} {
{m_{\nu}\upsilon_0^2}} \left (\frac{G_F}{2\sqrt{2}} \right )^2
F(X),
\\ F(X)&=& \frac{2}{\sqrt{\pi}}e^{-(1+X^2)}\sinh2X,\, \quad
X=\frac{1}{\upsilon_0}\sqrt{ 2\left (\frac{
\epsilon}{m_{\nu}}-1\right )} \earr
 As expected the
cross section exhibits resonance behavior though the normalized
function F(X) as shown in Fig. \ref{fig:ecapwidth}.
It is, of course, more practical to exhibit the function $F(X)$ as
a function of the the energy $\epsilon$. This is exhibited in
Fig.\ref{fig:ecapwidthE}. From this figure we see that the cross
section resonance is quite narrow. We find that the maximum occurs
at $\epsilon=m_{\nu}\left (1+2.8 \times 10^{-7}\right
)=50\mbox{keV+0.014eV}$ and has a width
$\Gamma=m_{\nu}(1+9.1\times 10^{-7})-m_{\nu}(1+0.32\times
10^{-7})\approx 0.04$eV. So for all practical purposes it is a
line centered at the neutrino mass. The width may be of some
relevance in the special case whereby the excited state can be
determined by atomic de-excitations at the sub eV level, but it
will not show up in the nuclear de-excitations.

If there is a resonance in the final nucleus at the energy
$E_x=\epsilon+(\Delta-b)$ with a width $\Gamma$ then perhaps it can be
 reached even if  $\epsilon$ is a bit larger than $m_{\nu}$,
e.g. $\epsilon=m_{\nu}+\Gamma/2$. The population of this resonance
can be determined by measuring the energy of the de-excitation
$\gamma$-ray, which should exceed by $\epsilon$ the maximum
observed in ordinary electron capture.

For antineutrinos having zero kinetic energy   the atom in the
final state has to have an excess energy $\Delta-(b-m_\nu)$ and
this can only happen if this energy can be radiated out via photon
or phonon emission. The photon emission takes place either as
atomic electron or nuclear level transitions. In the first case
photon energies falling in the eV-keV energy region and this
implies that only nuclei with a very small $\Delta$-value could be
suitable for this detection. In the second case, there should
exist a nuclear level that matches the energy difference
$E_x=\Delta-(b-m_\nu)$ and therefore   the incoming antineutrino
has no energy threshold. Moreover, spontaneous electron capture
decay is energetically forbidden, since this is allowed  for
$E_x<\Delta-(b+m_\nu)$.

As an example we consider the capture of a very low energy
$\bar\nu$ by the $^{157}_{65}$Tb  nucleus \beq \label{Tb1}
\bar\nu+ e^- +^{157}_{65}Tb \rightarrow ^{165}_{64}Gd^{*} \eeq
taking the allowed transitions from the ground state ($3/2^+$) of
parent nucleus,  $^{157}_{65}$Tb,  to the first excited $ 5/2^+$
state of the daughter nucleus $^{157}_{64}$Gd. The spin and parity
of the nuclei involved obey the relations $\Delta J=1$,
$\Pi_f\Pi_i=+1$, and the transition is dubbed as allowed. The
nuclear matrix element $ME$   can be written as can be written as
 \begin{equation}
|ME|^2 =   (g_{\scriptscriptstyle A}/g_{\scriptscriptstyle V})^2
\langle {\bf GT} \rangle^2
\end{equation}
where $g_{\scriptscriptstyle A}=1.2695$ and $g_{\scriptscriptstyle
V}=1$ being the axial and vector coupling constants respectively.
The nuclear matrix element is calculated using the microscopic
quasi-particle-phonon (MQPM) model~\cite{Toiv1,Toiv2} and it is
found to be $|ME|^2=0.96$. The   experimental value of first
excited $5/2^+$ is at 64~keV~\cite{Helmer} while the predicted by
the model   at 65~keV. The $\Delta$-value is  ranging from 60 to
63 keV~\cite{Helmer}.

 For K-shell electron
capture where  $\langle\phi_e\rangle^2=\left (\frac{\alpha
Z}{\pi}m_e\right )^3$ (1s capture) with binding energy
$b=50.24$keV, the velocity averaged  cross section takes the value
$$ \langle \sigma \upsilon \rangle =8.98\times 10^{-46}C^2_{\nu}
\hspace{2pt} \mbox{cm}^2$$ and  the event rate  we expect for mass
$m_t=1$kg is \beq R= 8.98\times 10^{-46} C^2_{\nu}
\hspace{2pt}\times 6\hspace{2pt}10^3\times
6.023\hspace{2pt}10^{23}\times \frac{m_t}{A\hspace{2pt}
}\hspace{2pt} \times 9.28\times 10^{17} \mbox{y}^{-1}=19
C^2_{\nu}\hspace{2pt}
 \mbox{y}^{-1} \eeq

    The life time of the source should be suitable for the experiment to be
performed. If it is too short, the time available will not be adequate for
the execution of the experiment. If it is too long, the number of counts
during the data taking will be too small. Then one will face formidable
backgrounds and/or large experimental uncertainties.

The source should be cheaply available in large quantities. Clearly a
compromise has to be made in the selection of the source. One can be optimistic
that adequate such quantities can be produced in Russian reactors. The nuclide parameters relevant to our work can be found in
\cite{Filianin14} (see also \cite{JDVNov10}), summarized in table \ref{taba}.
\begin{table}[htbp]
\begin{center}
\caption{ Nuclides with relevant for the search for the keV
sterile neutrinos in the electron capture process. We give the
life time $T_{1/2}$, the $Q$-value, the electron binding energy
$B_i$ for various captures and the value of $\Delta=Q-B_i$. For
details see ref. \cite{Filianin14}. \label{taba}}
\begin{tabular}{|c|c|c|c|c|c|c|c|}
\hline
Nuclide& $T_{1/2}$&
EC transition& $Q$ (keV)& $B_i$ (keV)&$B_j$ (keV)&$Q-B_i$
(keV)\\
\hline
$^{157}$Tb &71 y &3/2$^{+}\rightarrow 3/2^{-}$& 60.04(30)& K: 50.2391(5)& LI: 8.3756(5)& 9.76\\
$^{163}$Ho &4570 y &7/2$^{-}\rightarrow 5/2^{-}$& 2.555(16)& MI: 2.0468(5)& NI: 0.4163(5)& 0.51\\
$^{179}$Ta& 1.82 y &7/2$^+\rightarrow 9/2^+$& 105.6(4)& K: 65.3508(6)& LI: 11.2707(4) & 40.2\\
$^{193}$Pt& 50 y& 1/2$^-\rightarrow 3/2^+$& 56.63(30)& LI: 13.4185(3)& MI: 3.1737(17) & 43.2\\
$^{202}$Pb& 52 ky &0$^+\rightarrow 2 ^-$& 46(14)& LI: 15.3467(4)& MI: 3.7041(4) & 30.7\\
$^{205}$Pb& 13 My& 5/2$^-\rightarrow 1/2^+$& 50.6(5)& LI: 15.3467(4)& MI: 3.7041(4) & 35.3\\
$^{235}$Np &396 d& 5/2$^+\rightarrow 7/2^-$& 124.2(9)& K: 115.6061(16)& LI: 21.7574(3) &8.6\\
\hline
\end{tabular}
\end{center}
\end{table}
\section{Modification of the end point spectra of $\beta$ decaying nuclei}

The end point spectra of $\beta$ decaying nuclei can be modified
by the reaction involving sterile (anti)neutrinos \beq \nu+A(N,Z)
\rightarrow A(N-1,Z+1)+e^{-} \eeq or \beq \bar\nu+A(N,Z)
\rightarrow A(N+1,Z-1)+e^{+} \eeq

 This can be exploited in on ongoing  experiments, e.g. in
the Tritium decay \beq \label{tritium} \nu+^{3}_{1}H \rightarrow
^{3}_{2}He+e^{-} \eeq

 The relevant cross section is: \beq \sigma(E_{\nu})=
C^2_{\nu}\frac{1}{\upsilon} |ME(E_x)|^2_{\mbox{\tiny nuc}} \left
(\frac{G_F}{2\sqrt{2}} \right )^2
\frac{d^3p_A}{(2\pi)^3}\frac{d^3p_e}{(2\pi)^3}(2 \pi)^4\delta({\bf
p}_{\nu}-{\bf p_A-p_e}) \delta(E_{\nu}+\Delta-E_{e}) \eeq where
$\Delta$ is the atomic mass difference. Integrating over the
nuclear recoil momentum  and the direction of the electron
momentum we get:
 \beq
\sigma(E_{\nu})= C^2_{\nu}\frac{1}{\upsilon}
|ME(E_x)|^2_{\mbox{\tiny nuc}} \left (\frac{G_F}{2\sqrt{2}} \right
)^2 \frac{1}{\pi}  E_e P_e
 \eeq
where \beq \label{v1}
E_e=m_{\nu}+\frac{1}{2}m_{\nu}\upsilon^2+\Delta +m_e
 \eeq
 and
 \beq P_e= \sqrt{E_e^2-m_e^2}
 \eeq

\begin{figure}[htb]
  \begin{center}
\includegraphics[scale=1,width=8cm]{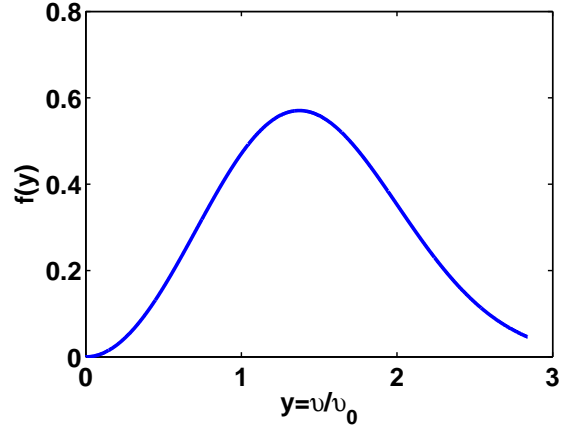}
 \caption{ The shape of f(y) for  the decay of $^{3}H$, where the atomic mass difference
 between$^{3}H$ and $^{3}He$ is taking    $\Delta=18.591 \mbox{keV}$~\cite{Audi02}. }
 \label{fig:nuescatd}
 \end{center}
  \end{figure}

 Folding the cross section with
the velocity distribution we find \beq \langle \sigma \upsilon
\rangle =C^2_{\nu} \left (\frac{G_F}{2\sqrt{2}} \right
)^2\hspace{2pt}  \frac{2}{\pi^{3/2}}\int_{0}^{y_{esc}} dy f(y)
 \eeq
 where
\beq f(y)=|ME|^2\hspace{2pt} y
\hspace{2pt}\mbox{sinh}(2y)\hspace{2pt} E_e P_e
e^{-(1+y^2)}F(Z_f,E_e) \eeq
 with $$y=\upsilon/\upsilon_0$$
 The
Fermi function, $F(Z_f,E_e)$ encapsulates the effects of the
Coulomb interaction for a given lepton energy $E_e$ and final
state proton number $Z_f$. The function $f(y)$ is exhibited in Fig. \ref{fig:nuescatd}.

In   transitions happening inside the same isospin multiplet
($J^\pi \rightarrow J^\pi, J \ne 0$) both the vector and axial
form factors contribute and in this case the nuclear matrix
element $ME(E_x)$   can be written as can be written as
\begin{equation}
|ME|^2 =  \langle {\bf F} \rangle^2 + (g_{\scriptscriptstyle
A}/g_{\scriptscriptstyle V})^2 \langle {\bf GT} \rangle^2 \, ,
\label{allowed}
\end{equation}
where $g_{\scriptscriptstyle A}=1.2695$ and $g_{\scriptscriptstyle
V}=1$ being the axial and vector coupling constants respectively.
In case of $^{3}H$ target we adopt $\langle {\bf F} \rangle^2 =
0.9987$ and $\langle {\bf GT} \rangle^2 = 2.788$
from~\cite{Schiavilla98}. Thus $|ME|^2=5.49$.

Thus the velocity averaged  cross section takes the value
$$\langle \sigma \upsilon
\rangle =3.44\times 10^{-46}C^2_{\nu} \hspace{2pt} cm^2$$
  and  the expected event  rate becomes
\beq R= 3.44\times 10^{-46}C^2_{\nu}\hspace{2pt}\times
6\hspace{2pt}10^3\times 6.023\hspace{2pt}10^{23}\times
\frac{m_t}{A\hspace{2pt} }\hspace{2pt} \times 9.28\times 10^{17}
\mbox{y}^{-1}
\eeq
For a  mass of the current KATRIN target, i.e. about 1 gr, we get
\beq
 R= 0.380 C^2_{\nu} \hspace{2pt}
 \mbox{y}^{-1} \eeq

 It is interesting to compare the neutrino capture rate
 \begin{equation}\label{app:Kend1}R_\nu=\langle \sigma \upsilon
\rangle \frac{\rho}{m_{\nu}}=3.44\times
10^{-46}C^2_{\nu}\hspace{2pt}\times 6\hspace{2pt}10^3\times
9.28\times 10^{17}=1.91\times 10^{-24}C^2_{\nu}\hspace{2pt}y^{-1}
\end{equation}
 with that of beta decay process $\tritium \to \Hethree + e^{-} + \overline{\nu}_j$
,  whose rate $R_\beta$ is given by
\begin{equation}
\label{app:Kend2} R_\beta = \frac{G_F^2}{2 \pi^3 }\hspace{2pt}
\int_{m_e}^{W_o} p_e E_e F(Z,E_e) |ME|^2 E_\nu p_\nu \, d E_e
\label{ratedecay}
\end{equation}
where ${W_o}$ is the maximal electron energy or else beta decay
endpoint
\begin{equation}
\label{app:Kend3}
{W_o}= K_{end} +m_e
\end{equation}
with
\begin{equation}
\label{app:Kend4}
K_{end}= \frac{(m_{\tritium}-m_e)^2 -
(m_{\Hethree}+m_{\nu})^2}{2m_{\tritium}}\simeq\Delta=18.591\hspace{2pt}
\mbox{keV}
\end{equation}
the electron kinetic energy at the endpoint, and
\begin{eqnarray}
   & m_{e} \approx 510.998 910(13) \hspace{2pt}keV \\
   & m_{\tritium}  \approx 2808920.8205 (23) \hspace{2pt}keV \\
   & m_{\Hethree}  \approx  2808391.2193 (24) \hspace{2pt}keV
\end{eqnarray}
 Masses $m_{\tritium}$ and $m_{\Hethree}$ are nuclear masses~\cite{Audi02,RevModPhys67,Beringer:2012zz}.
The calculation of (\ref{app:Kend2}) gives $ R_\beta =0.055
y^{-1}$. The ratio of $R_\nu$ to corresponding beta decay
$R_\beta$ is very small.
\begin{equation}
R_\nu = 0.034 \cdot 10^{-21}C^2_{\nu}\hspace{2pt}R_\beta
\end{equation}
The situation is more optimistic  in a narrow interval
$W_o-\delta<E_e<W_o$ near the endpoint. As an example, we consider
an energy resolution $\delta=0.2$~eV close to the  expected
sensitivity of the KATRIN experiment~\cite{katrin}. Then the ratio
of the event rate $R_\beta(\delta=0.2eV)$  to that of neutrino
capture $R_\nu$ gives
$$R_\nu  = 5.75 \cdot 10^{-9}C^2_{\nu}\hspace{2pt}R_\beta(\delta=0.2eV)$$
 \begin{figure}
  \begin{center}
\includegraphics[scale=1,width=0.5\textwidth]{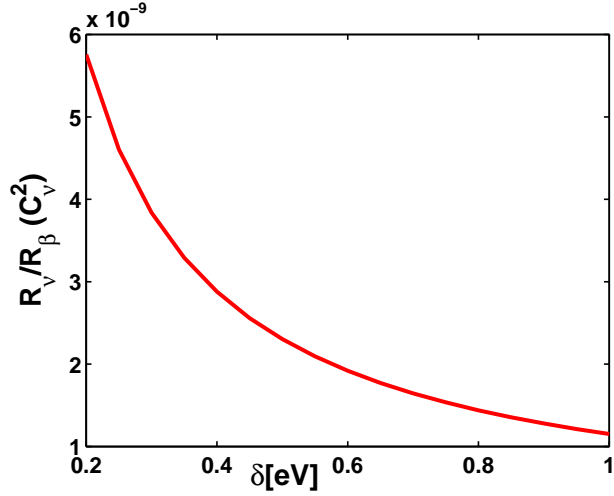}\\
 \caption{Ratio of decay rates $R_\nu/R_\beta$ (in units of $C^2_{\nu}$) as a function of energy resolution $\delta$ near the endpoint.
 \label{resolution}}
 \end{center}
  \end{figure}
In fig.~\ref{resolution} we present the ratio of the event rate
decay rate of $R_\beta(\delta)$ for the beta decay compared with
the neutrino capture rate $R_\nu$ as a function of the energy
resolution $\delta$ in the energy region $W_o-\delta<E_e<W_o$

 Moreover,  the electron   kinetic energy $K_e$  due to
neutrino capture process (\ref{tritium}) is
\begin{equation}\label{Ke}
 K_e= E_{\nu}+K_{end}>m_{\nu}\hspace{2pt}+18.591\hspace{2pt}
\mbox{keV}
\end{equation}
this means that the electron in the final state has a kinetic
energy of at least $m_{\nu}$ above the corresponding beta decay
endpoint energy. There is no reaction induced background there, but, unfortunately, the ratio obtained above is much lower than the expected KATRIN sensitivity.

\section{Discussion}

In the present paper we examined the possibility of direct
detection of sterile neutrinos of a mass 50 keV, in dark matter
searches. This depends on finding solutions to two problems. The
first is the amount of energy expected to be deposited in the
detector and the second one is the expected event rate.
 In connection with the energy  we have seen that, even though these neutrinos are quite
heavy, their detection is not easy, since like all dark matters
candidates move in our galaxies with not relativistic velocities,
  $10^{-3}$c on the  average and  with energies
about 0.05 eV, not all of which can be deposited in the detectors.
Thus the detection techniques employed in the standard dark matter
experiments, like those looking for  heavy WIMP candidates, are
not applicable in this case.

We started our investigation by considering neutrino electron
scattering. Since the energy of the sterile neutrino is very small
one may have to consider systems with very small electron binding,
e.g. electron pairs in superconductors, which are limited to
rather small  number of electron pairs. Alternatively one may use
low temperature bolometers, which can be larger in size resulting
in a higher expected event rate. These experiments must  be able
to detect very small amount of energy.

Then we examined more exotic options by exploiting atomic and
nuclear physics. In atomic physics we examined the possibility of
spin induced excitations. Again to avoid background problems the
detector has to a crystal operating at low temperatures. Then what
matters is the atomic structure of the ions of the crystal or of
suitably implanted impurities. The rate in this case is less than
that obtained in the case of bolometers, but one may be able to
exploit the characteristic feature of the spectrum of the emitted
photons.

From  the nuclear physics point of view,  we consider
  the antineutrino  absorption on an electron
capturing  nuclear system leading to a fine resonance in the
(N+1,Z-1) system, centered 50 keV above the highest excited state
reached by the ordinary electron capture. The de-excitation of
this resonance will lead to a very characteristic $\gamma$ ray.
Finally the sterile neutrino will lead to $\nu
+$A(N,Z)$\rightarrow e^{-}$+A(N-1,Z+1) reaction. The produced
electrons will have  a maximum  energy which goes beyond the end
point energy of the corresponding $\beta$ decay by essentially the
neutrino mass. The signature is less profound than in the case of
antineutrino absorption.

Regarding the event rate, as we have mentioned before, it is
proportional to the coupling of the sterile neutrino to the usual
electron neutrino indicated above as  $C_{\nu}^2$. This parameter
is not known. In neutrino  oscillation experiments a value of
$C_{\nu}^2 \approx 10^{-2}$ has been employed. With such a value
our results show that the 50 keV neutrino is detectable in the
experiments discussed above. This large value of $C_{\nu}^2$ is
not consistent, however,  with  a sterile  50 kev neutrino. In
fact such a neutrino  would have a life time \cite{DolgovHansen02}
of $2 \times 10^{5}$y, much shorter than the age of the universe.
A cosmologically viable sterile 50 keV neutrino is allowed to
couple to the electron neutrino  with  coupling $C_{\nu}^2<
1.3\times 10^{-7}$. Our calculations indicate that such a neutrino
is not directly detectable with experiments considered in this
work. The  results, however, obtained  for the various physical  processes considered in this work, can be very useful in the analysis of the possible  experimental searches of lighter sterile neutrinos in the mass range of 1-10 keV.



{\bf Acknowledgements}: The authors are indebted to professor  Marco Bernasconi
for useful suggestions in connection with the phonon excitations of low temperature bolometers.

\clearpage

\section{Appendix A: Angular momentum coefficients entering atomic excitations.}
The angular momentum coefficients entering single particle
transitions are shown in tables \ref{tab:tab1}-\ref{tab:tab2}.
\begin{table}[htb]
\caption{the coefficients $\left
(C_{\ell,j_1,m_1,j_2,m_2}\right)^2 $ connecting via the spin
operator a given initial state
 $|i\rangle=|n\ell,j_1,m_1\rangle$ with all possible states $|f \rangle=|n\ell,j_2,m_2\rangle $, for $\ell=0,\,1$. Note s-states are favored.}
\renewcommand{\arraystretch}{1.5} 
\renewcommand{\tabcolsep}{0.2pc} 
\label{tab:tab1}
$$
\left(
\begin{array}{ccccc|c}
\ell&j_1&m_1&j_2&m_2&C^2_{\ell,j_1,m_1,j_2,m_2}\\
\hline
 0 & \frac{1}{2} & -\frac{1}{2} & \frac{1}{2} & \frac{1}{2} & 2 \\
\end{array}
\right),
\left (
\begin{array}{ccccc|c}
&|i\rangle&&|f\rangle&&\\
\hline
\ell&j_1&m_1&j_2&m_2&C^2_{\ell,j_1,m_1,j_2,m_2}\\
\hline
 1 & \frac{1}{2} & -\frac{1}{2} & \frac{1}{2} & \frac{1}{2} &
   \frac{2}{9} \\
 1 & \frac{1}{2} & -\frac{1}{2} & \frac{3}{2} & -\frac{3}{2} &
   \frac{4}{3} \\
 1 & \frac{1}{2} & -\frac{1}{2} & \frac{3}{2} & -\frac{1}{2} &
   \frac{8}{9} \\
 1 & \frac{1}{2} & -\frac{1}{2} & \frac{3}{2} & \frac{1}{2} &
   \frac{4}{9} \\
 1 & \frac{1}{2} & \frac{1}{2} & \frac{3}{2} & -\frac{1}{2} &
   \frac{4}{9} \\
 1 & \frac{1}{2} & \frac{1}{2} & \frac{3}{2} & \frac{1}{2} &
   \frac{8}{9} \\
 1 & \frac{1}{2} & \frac{1}{2} & \frac{3}{2} & \frac{3}{2} & \frac{4}{3}
   \\
 1 & \frac{3}{2} & -\frac{3}{2} & \frac{3}{2} & -\frac{1}{2} &
   \frac{2}{3} \\
 1 & \frac{3}{2} & -\frac{1}{2} & \frac{3}{2} & \frac{1}{2} &
   \frac{8}{9} \\
 1 & \frac{3}{2} & \frac{1}{2} & \frac{3}{2} & \frac{3}{2} & \frac{2}{3}
   \\
\end{array}
\right)
$$
\end{table}

\begin{table}[htb]
\caption{The same as in table \ref{tab:tab1}, the coefficients
$\left (C_{\ell,j_1,m_1,j_2,m_2}\right )^2$ for $\ell=2$}
\renewcommand{\arraystretch}{1.5} 
\renewcommand{\tabcolsep}{0.2pc} 
\label{tab:tab2}
$$
\left(
\begin{array}{ccccc|c}
&|i\rangle&&|f\rangle&&\\
\hline
\ell&j_1&m_1&j_2&m_2&C^2_{\ell,j_1,m_1,j_2,m_2}\\
\hline
 2 & \frac{3}{2} & -\frac{3}{2} & \frac{3}{2} & -\frac{1}{2} &
   \frac{6}{25} \\
 2 & \frac{3}{2} & -\frac{3}{2} & \frac{5}{2} & -\frac{5}{2} &
   \frac{8}{5} \\
 2 & \frac{3}{2} & -\frac{3}{2} & \frac{5}{2} & -\frac{3}{2} &
   \frac{16}{25} \\
 2 & \frac{3}{2} & -\frac{3}{2} & \frac{5}{2} & -\frac{1}{2} &
   \frac{4}{25} \\
 2 & \frac{3}{2} & -\frac{1}{2} & \frac{3}{2} & \frac{1}{2} &
   \frac{8}{25} \\
 2 & \frac{3}{2} & -\frac{1}{2} & \frac{5}{2} & -\frac{3}{2} &
   \frac{24}{25} \\
 2 & \frac{3}{2} & -\frac{1}{2} & \frac{5}{2} & -\frac{1}{2} &
   \frac{24}{25} \\
 2 & \frac{3}{2} & -\frac{1}{2} & \frac{5}{2} & \frac{1}{2} &
   \frac{12}{25} \\
 2 & \frac{3}{2} & \frac{1}{2} & \frac{3}{2} & \frac{3}{2} &
   \frac{6}{25} \\
 2 & \frac{3}{2} & \frac{1}{2} & \frac{5}{2} & -\frac{1}{2} &
   \frac{12}{25} \\
 2 & \frac{3}{2} & \frac{1}{2} & \frac{5}{2} & \frac{1}{2} &
   \frac{24}{25} \\
 2 & \frac{3}{2} & \frac{1}{2} & \frac{5}{2} & \frac{3}{2} &
   \frac{24}{25} \\
 2 & \frac{3}{2} & \frac{3}{2} & \frac{5}{2} & \frac{1}{2} &
   \frac{4}{25} \\
 2 & \frac{3}{2} & \frac{3}{2} & \frac{5}{2} & \frac{3}{2} &
   \frac{16}{25} \\
 2 & \frac{3}{2} & \frac{3}{2} & \frac{5}{2} & \frac{5}{2} &
   \frac{8}{5} \\
 2 & \frac{5}{2} & -\frac{5}{2} & \frac{5}{2} & -\frac{3}{2} &
   \frac{2}{5} \\
 2 & \frac{5}{2} & -\frac{3}{2} & \frac{5}{2} & -\frac{1}{2} &
   \frac{16}{25} \\
 2 & \frac{5}{2} & -\frac{1}{2} & \frac{5}{2} & \frac{1}{2} &
   \frac{18}{25} \\
 2 & \frac{5}{2} & \frac{1}{2} & \frac{5}{2} & \frac{3}{2} &
   \frac{16}{25} \\
 2 & \frac{5}{2} & \frac{3}{2} & \frac{5}{2} & \frac{5}{2} & \frac{2}{5}
   \\
\end{array}
\right)
 $$
\end{table}

\section{Appendix B: Exotic atomic experiments}

As we have mentioned the atomic experiment has to be done at low
temperatures. It may be difficult to find materials exhibiting
atomic structure at low temperatures, It amusing to note that one
may be able to employ at low temperatures some exotic materials
used in quantum technologies (for a recent review see
\cite{NVReview13}) like nitrogen-vacancy (NV), i.e. materials
characterized by spin $S=1$, which in a magnetic field  allow
transitions between $m=0,\,m=1$ and $m=-1$. These states are spin
symmetric. Antisymmetry requires the space part to be
antisymmetric, i.e. a wave function of the form
$$\psi=\phi^2_{n\ell}(r)\left
[L=\mbox{odd,}S=1\right]\hspace{2pt} J=L-1,L,L+1$$ Of special
interest are:
$$\psi=\phi^2_{n\ell}(r)^3P_J, \hspace{4pt}\phi^2_{n\ell}(r)^3F_J$$
Then the spin matrix element takes the form:
$$\langle ^3L_{J_2m_2}|\sigma|^3L_{J_1m_1}\rangle=\frac{1}{\sqrt{2J_2+1}}\langle J_1m_1,1m_2-m_1|J_2m_2\rangle
\langle^3L_{J_2}||\sigma||^3L_{J_1}\rangle,\,L=P,F$$ The reduced
matrix elements are given in table \ref{tab:tab3}, as well as  the
full matrix element $\langle
^3P_{J_2m_2}|\sigma|^3P_{J_1m_1}\rangle^2$ of the most important
component.

\begin{table}
\caption{The coefficients $
\langle^3P_{J_2}||\sigma||^3P_{J_1}\rangle$,  $
\langle^3F_{J_2}||\sigma||^3F_{J_1}\rangle$ and
$\langle^3P_{J_2m_2}|\sigma|^3P_{J_1m_1}\rangle^2 $. For the
notation see text.}
\renewcommand{\arraystretch}{1.5} 
\renewcommand{\tabcolsep}{0.2pc} 
 \label{tab:tab3}
$$
\begin{array}{cc|c}
J_1&J_2&\langle^3P_{J_2}||\sigma||^3P_{J_1}\rangle\\
\hline
 0 & 1& \sqrt{\frac{2}{3}}\\
 1 & 1& \frac{1}{\sqrt{2}}\\
1 & 2& \sqrt{\frac{5}{6}}\\
2 & 2& \sqrt{\frac{5}{2}}\\
\end{array}
\hspace{10pt}
\begin{array}{cc|c}
J_1&J_2&\langle^3F_{J_2}||\sigma||^3F_{J_1}\rangle\\
\hline
 2 & 2& -\frac{\sqrt{10}}{3}\\
 2 & 3& \frac{2\sqrt{5}}{3}\\
3 & 3& \frac{\sqrt{7}}{6}\\
 3 & 4& \frac{3}{2}\\
 4 & 4& \frac{\sqrt{15}}{2}\\
\end{array}
\hspace{10pt}
\begin{array}{cccc|c}
J_1&m_1&J_2&m_2&\langle^3P_{J_2m_2}|\sigma|^3P_{J_1m_1}\rangle^2\\
\hline
 0 & 0&1&m_2& \frac{2}{9}\\
 1 & -1& 1&-1&\frac{1}{6}\\
1 & -1& 1&0&\frac{1}{6}\\
 1 & 0& 1&0&0\\
 1 & 0& 1&1&\frac{1}{6}\\
 1 & 1& 1&1&\frac{1}{6}\\
\end{array}
$$
\end{table}
\clearpage
\end{document}